# Sustained unidirectional rotation of a self-organized DNA rotor on a nanopore

Xin Shi,[1] Anna-Katharina Pumm,[2] Jonas Isensee,[3] Wenxuan Zhao,[1] Daniel Verschueren,[1, †] Alejandro Martin-Gonzalez,[1] Ramin Golestanian,[3,4,*] Hendrik Dietz,[2,*] Cees Dekker[1,*]

[1] Department of Bionanoscience, Kavli Institute of Nanoscience Delft, Delft University of Technology, Delft, The Netherlands.
[2] Department of Physics, Technical University of Munich, Garching 85748, Germany
[3] Max Planck Institute for Dynamics and Self-Organization, Göttingen 37077, Germany
[4] Rudolf Peierls Centre for Theoretical Physics, University of Oxford, OX1 3PU, Oxford, UK
† Current address: The SW7 Group, 86/87 Campden Street, W8 7EN, London, UK
* Corresponding authors. Emails: c.dekker@tudelft.nl, dietz@tum.de, ramin.golestanian@ds.mpg.de

**Abstract**

Flow-driven rotary motors drive functional processes in human society such as windmills and water wheels. Although examples of such rotary motors also feature prominently in cell biology, their synthetic construction at the nanoscale has thus far remained elusive. Here, we demonstrate flow-driven rotary motion of a self-organized DNA nanostructure that is docked onto a nanopore in a thin solid-state membrane. An elastic DNA bundle self assembles into a chiral conformation upon phoretic docking onto the solid-state nanopore, and subsequently displays a sustained unidirectional rotary motion of up to 20 revolutions/s. The rotors harness energy from a nanoscale water and ion flow that is generated by a static (electro)chemical potential gradient in the nanopore that is established through a salt gradient or applied voltage. These artificial nanoengines self-organize and operate autonomously in physiological conditions, paving a new direction in constructing energy-transducing motors at nanoscale interfaces.

**Main Text**

Nanoscale rotary motors like $F_oF_1$-ATP synthase [1,2] or the flagellar motors [3], are vital to cellular functionality as they drive the fundamental biological processes of ATP synthesis and bacterial propulsion, respectively. These exemplary motor proteins self-assemble into nanoscale thin[4] membranes that span cellular compartments, and convert energy from a gradient-induced local ion flow into mechanical rotary motion. Building synthetic rotary nanomachines that work at such nanoscale interfaces is a beckoning perspective and a critical step towards building complex artificial machinery that can operate in the out-of-equilibrium conditions present in cells [5-7].

Autonomous sustained directed motion of a nanomachine is a manifestation of its ability to transduce energy and perform work. In contrast to macroscale machines, nanoscale structures are subject to thermal fluctuations which cause random Brownian motion of their moveable parts [8], and no sustained directed motion can emerge from these fluctuations without the consumption of free energy [9,10]. Chemists have built a large variety of dynamic nanostructures whose motion can be regulated, e.g. molecular switches [7], mechanically interlocking molecules [11], and self-assembled DNA structures [5]. So far, however, these objects do not exhibit an autonomously sustained unidirectional operation, as dynamics in these systems could only be induced by the deliberate alteration of external experimental conditions such as the addition of chemical compounds [7], synchronized exposure to light [12], or the timely alternation of external electric fields [13]. Modulation steps that are manually applied by the external operator then move these nanomachines akin to puppets on a string, sequentially pushing them into a new equilibrium state at every intervention. DNA walkers can perform tasks such as cargo sorting through DNA-strand-displacement reactions [14], but do not autonomously transduce and consume energy in a cyclic way. And whereas biological rotary molecular engines such as ATP synthase can be repurposed to drive dynamics in hybrid nanostructures [15,16], the reverse – establishing purely synthetic rotary nanomachines that consume free energy to perform mechanical work – have so far remained challenging. [17]

Here, we demonstrate autonomous and continuous unidirectional rotation of an electrophoretically self-configured chiral DNA nanostructure that is driven by the flux of water and ions through a nanoscale opening in a thin solid-state membrane. Such a nanoscale flux can be induced from physiologically relevant energy sources by simply applying a static voltage or a salt gradient across the solid-state membrane. After docking the nanostructure onto a nanopore, it rotates unidirectionally, a process that can be monitored in real time by fluorescence microscopy and sub-diffraction-limit single-particle tracking. Supported by discrete-elastic-rod modelling, we deduce that this motion results from a field-induced shape deformation of the DNA bundle, that is driven into rotary motion by the water flux through the nanopore.

**Design of the DNA rotary machine on a nanopore**

Rotary nanomachines were constructed from DNA rods that self-configured after being docked onto a nanopore. Each rotor was composed of a T-shaped DNA structure (Fig. 1) that primarily consisted of a 450-nm long DNA 6-helix origami bundle (6hb) configured in a honeycomb lattice [18]. A 50-nm dsDNA protruding from its center point was included for biasing the docking of the bundle onto the nanopore towards its center and for increasing the amount of negative charge inside the nanopore to enhance the induced electroosmotic flow. For optical monitoring, a group of 20 cyanine fluorescent dyes was included at the end of one tip (or both tips) of the 6hb. Details of the design and folding of DNA structure are included in Fig. S1 and the Methods section. Folded structures were characterized with Atomic Force Microscopy which showed long bundles with an average end-to-end distance of about 410 nm (Fig. 1b and Fig. S2).

These DNA rods were docked onto a 20x20 array of ~50-nm-diameter solid-state nanopores in a 20-nm thin silicon-nitride (SiN$_x$) membrane (Fig. 1, c and d). *Cis* and *trans* compartments were filled with the same buffer (50 mM NaCl) and a transmembrane voltage was applied (100 mV), or alternatively, they were filled with a buffer of different ionic concentration (typically *cis*:*trans* 50 mM:550 mM NaCl) with no transmembrane voltage applied. In both these two experimental conditions, the DNA nanostructures could be docked onto the nanopore by, respectively, electrophoresis [19] or diffusiophoresis [20,21] (Fig. 1e and Fig. S3). The structural rigidity of the 450-nm-long 6hb, which has a persistence length of about ~1.5 μm [22], allowed for permanent docking when the beam docked parallel to the membrane surface, since its rigidity prevented translocation of the structure through the 50-nm nanopore. However, DNA rods that entered the pore with the tip of the bundle translocated to the *trans* side and were not further studied. Since both the DNA structure and the SiN$_x$ surface are highly negatively charged and hydrophilic, the docked DNA rotors were found to move relatively freely on the surface.

**Unidirectional rotation of DNA nanostructure driven by a voltage-induced flow**

The DNA rotors were found to exhibit a sustained unidirectional rotation that was induced by an externally applied transmembrane DC voltage (Fig. 2a). Figure 2b shows a series of snapshots of a single fluorescent spot for a docked DNA structure on a nanopore, which is illustrative of a rotary motion performed by the nanostructure. The images reveal a clear directional rotation of the spot around the pore (red cross at center). Figure 2c shows a heatmap of sub-pixel tip locations for this rotor, which clearly shows a circular trajectory of the tip. An example trajectory of 20 consecutive frames of the tip of the nanorod is overlaid which displays the unidirectional clockwise motion of this DNA rotor. The cumulative angular displacement $\theta(t)$ of the tip (Fig. 2d, corresponding step size distribution see Fig. S4) shows that this rotary motion was maintained for hundreds of rotations in the period of recording (40 seconds in this case). Continuous directional rotation of the DNA structure is evident from the downward sloped line that displays ~5 revolutions per second, which contrasts Brownian dynamics that would produce random angular fluctuations. Figure 2e displays cumulative angular displacements $\theta(t)$ of 40 DNA rotors from a single experiment, where the linear curves indicate the clearly driven motion of the nanorods on the nanopores. The directed motion is further illustrated by the superlinear mean-square displacement (MSD) curves of these rotors (Fig. 2f), which starkly contrast a linear MSD curve that would result from mere Brownian motion.

The DNA rotors displayed considerable variations in rotation direction and speed. The dynamics for different docked rotors ranged from unbiased Brownian motion to driven rotation with up to ~20 revolutions per second, while within a single trajectory, the speed and direction of rotation was generally well maintained. Figure panels 2g-i show selected typical examples of the heterogeneous dynamics of DNA structures that differed from the clear driven rotation of Fig. 2c., viz., examples of rare directional switching (<1% of all docked structures, Fig. 2g), unbiased Brownian motion (Fig. 2h), and stick-slip behavior (Fig. 2i). We attribute this heterogeneity to the self-configuring nature of the DNA rotors that, upon docking onto the pore, results in a variety of configurations of the DNA structure, as will be discussed later in the simulation results. Analyzing typical experiments (Fig. S5a), we found that 78% of rotors that displayed dynamics (see Data Analysis Section in Methods, Supplementary Information) showed driven rotation. The rotation speed of the rotors increased with increasing bias voltage. As shown in Fig. 2j, k, a switching of the transmembrane bias voltage from 50mV to 100mV led to a sudden increase in the rotation velocity at the point of switching (Fig. 2j). Upon increasing the applied voltage, the average speed of most rotors increased, while furthermore many stationary ones were put into rotary motion (Fig. 2k). We tested the stability of the rotational speed by stepping the bias voltage up and down after the rotation was established, see Supplementary Figure S6. These data showed that the rotational speed of the rotors typically (though not exclusively) reverted to a similar rotational speed after the voltage recovered the initial value.

Subsequently, to reveal their configurations, we labelled the two opposite tips of the 6hb with different colors (Cy3 and Cy5, Fig. 3a). Figure 3b-d shows typical examples of heatmaps of the tip locations of both ends of a rotary nanomachine (more examples in Fig. S7 and Supplementary Video 1). These data show that the end-to-end distances were between 50-350 nm (Fig. 3e), i.e., they typically were found to be much shorter than the full 450 nm length of the straight 6hb (420 nm if the span of fluorophore labels is taken into account, see Fig. S1). This reduced end-to-end distance was maintained during the rotation (Fig. S7) and indicates that the rod-shaped bundles were significantly mechanically deformed during rotation. A control experiment using elongated nanopores showed that DNA bundles that docked onto the nanopores stably self-aligned to the long axis of the pores indicating a slight wedging of rods into the pore (Fig. S8). Finally, we observed that the capture and bending did not necessarily happen at the center of the DNA bundle, but instead could occur off-center. As shown in Fig. 3c-d and Supplementary video 1, the radii of the circular trajectories of the two ends were often different.

From these experimental observations, we conclude that these elastic DNA nanorods on the nanopore self-adopt a stable bent conformation that breaks the structural mirror-symmetry inherent within the nanorod to sustain a persistent rotation. This self-organization must arise from the interplay between the nanopore and the charged elastic nanorod. Moreover, the spontaneous breaking of mirror symmetry, otherwise present in the system, provides a mechanism to torsionally couple the rotor to the electroosmotic flow that drives the rotation of the rotor. The fact that both clockwise (CW) and counterclockwise (CCW) rotations exist indicates that the introduced asymmetry is not intrinsic to the DNA structure itself, but rather an emergent feature resulting from this spontaneous symmetry breaking.

**Flow drives the electric-field-shaped DNA rotors**

To explore the symmetry breaking and torsional coupling, we simulated DNA rods docked on a nanopore using a simple discrete elastic-rod model [23] where the DNA 6hb was approximated as a chain of fixed-length rod segments. Given overdamped dynamics, we can describe the velocity $v$ of each rod segment by a local force balance (see Fig. 3h) between the internal constraint force ($\boldsymbol{F}_{\mathrm{con}}$), induced mechanical bending force ($\boldsymbol{F}_{\mathrm{bend}}$), the effective electrophoretic force ($\boldsymbol{F}_{\mathrm{ep}}$), the electroosmotic force ($\boldsymbol{F}_{\mathrm{eof}}$), and the viscous drag force $\boldsymbol{F}_{\mathrm{drag}} = \boldsymbol{Z}\boldsymbol{v}$ when motion is induced,

$$\boldsymbol{Z}\boldsymbol{v} + \boldsymbol{F}_{con} + \boldsymbol{F}_{bend} + \boldsymbol{F}_{ep} + \boldsymbol{F}_{eof} = \boldsymbol{0} \qquad \text{(eqn. 1)}.$$

Here, $\boldsymbol{Z}$ is the anisotropic friction tensor of the rod segment, which has coefficients $\xi_\parallel$ and $\xi_\perp$ along the principal axes parallel and perpendicular to the segment, respectively. Note that phoretic transport processes are generally force-free but can be represented by an equivalent force in this context. For the simulations, we adopted physically realistic parameters for the DNA rod stiffness and electrophoretic and electroosmotic forces (See Methods and Supplementary Note 2).

The simulations showed that the applied electric field, which is symmetrically radiating from the nanopore, instigates self-configuration by significantly deforming the 6hb into a stable chiral conformation. The electrophoretic forces pull the strongly negatively charged segments towards the pore, yielding a new shape where the stiffness of the rod counteracts the electrophoretic pull (Fig. 3f). The simulations showed considerable deformations of the 6hb for a wide range of effective electrophoretic force values that, due to counterion condensation and local electroosmotic flow induced by the DNA itself [19], are reduced from what the bare DNA charge density would dictate. The docked DNA rotors adopt a variety of stable chiral shapes, ranging from an archetypical S or Ƨ shape to U shapes and loop structures (Fig. 3i, i-iv, see Fig. S9 for their side views, Supplementary Movie 2-6 for additional examples). The self-configured shape that results upon docking was found to depend on the initial in-plane position of the DNA bundle relative to the nanopore, analogous to different approaching directions in the actual docking experiments. Centered initial positions produced S or Ƨ shapes and off-center initial positions yielded more asymmetric U shapes. Similar to the experiments, the simulated rotors were found to exhibit strongly reduced end-to-end distances, see Fig. 3j. These various shapes explain the experimental observations of different rotation directions (Fig. 2e), the wide range of reduced end-to-end distances (Fig. 3e), and the observation that some rotors rotate with their ends in tandem (Fig. 3d).

Next, we found that the water flow through the nanopore drives these self-configured chiral nanostructures into a persistent and constant rotational motion (Fig. 3k). The essential point is that the radial electroosmotic flow couples to the non-radially oriented segments of the deformed chiral 6hb (Fig. 3g), which, due to the friction coefficients $\xi_\parallel$ and $\xi_\perp$ that differ by a factor of about 2, leads to a nonvanishing net torque. As Fig. 3g shows, the electroosmotic drag is larger for DNA segments that are closer to the nanopore, indicating that the torque is primarily generated by the more central parts of the rotor whereas the further ends of the bundle chiefly act as a load. This is in line with the rapid decay of electric field strength away from the nanopore. As a result of the electroosmotic flow coupling, most S- and U-shaped rotors were found to be driven into a sustained rotation, while some structures either translocated through the pore or deformed mirror-symmetrically with respect to the pore, yielding a non-rotating state (Supplementary Movie 2). While the simulations provide a striking confirmation of many of the experimental results, the rotation speed in the simulation was much larger than experimentally observed, which may be attributed to a finite surface friction of the rotor on the membrane surface and to induced electroosmotic flow from the 6hb DNA itself, which both were not included in the model. The simulated rotors showed a weak correlation of end-to-end distance and rotation speed, with rotation speeds of simulated rotors peak around 200 nm (see Fig. S9). From a parameter scan for varying strengths of the electrophoresis and electroosmotic drag (Fig. S11), we found a wide parameter region where 6hbs rotated with various speeds, as well as neighboring regimes with non-rotating, translocating, and non-docking 6hbs. We also estimated the energy consumption rate of these rotors and obtained a result of about 190 pNnm/s (which corresponds to about 45 $k_BT$/s, see Supplementary Note 4).

**Unidirectional rotation of DNA nanomachines driven by a salt-gradient-induced flow**

We found that a simple salt gradient across the pore can self-organize and drive the DNA rotors as well, reminiscent of the transmembrane ion gradients in cells that are the primary energy source for rotary motor proteins such as the $F_o$ motor. Upon establishing a salt gradient of NaCl (*cis:trans* 50 mM:550 mM) across the $SiN_x$ membrane (and no external voltage applied), the nanomachines operated very similarly, again displaying sustained unidirectional rotations. Through diffusiophoresis, where a dissolved analyte exhibits an electrostatic interaction with ions [20,24], the negatively charged DNA rotors were pulled towards the nanopores from the low-salt towards the high-salt concentration, similar to the effect of the electric field under a transmembrane voltage. The same diffusiophoretic stress compresses the DNA nanorods and reconfigures the nanostructures into a chiral shape, and hence our simulations, which describe the rotors on a phenomenological level with effective forces, apply similarly. Upon docking of the DNA structures onto the nanopores, a water flow is established from *trans* to *cis*, a phenomenon known as diffusioosmosis, which is similar to the electroosmotic flow developed under a transmembrane voltage (Fig. 4a, Fig S13) [24][25].

Like the voltage-driven rotations, the ion-gradient-driven rotors exhibited a pronounced and sustained unidirectional rotation with up to ~15 revolutions/s. The observed behavior of the rotors was very similar to that observed under electrical bias conditions: a large portion of rotors (Fig. S4b) showed persistent unidirectional rotations at various speeds in either clockwise or counterclockwise directions. Fig. 4b and c show the cumulative angular displacements *θ(t)* over time and their corresponding MSD curves for 20 docked DNA structures. The linear angular-displacement curves and the superlinear MSD curves explicitly verify a clearly driven rotation of the structures, demonstrating that these nanomachines convert free energy from this static chemical gradient into mechanical work, which in our experiments was dissipated in friction with solvent.

**Concluding remarks**

Summing up, we demonstrated a self-organized nanoscale DNA rotary machine that is driven into a sustained unidirectional rotary motion by a nanofluidic flow generated in a nanopore through an electrochemical imbalance across a membrane. These nanomachines transduce free energy into work by exploiting physiologically relevant free-energy sources, viz., a static potential gradient that arises from simply maintaining two compartments at a different electrochemical potential across a nanometer-thin membrane. Operating autonomously without the need for external cyclical interventions by the operator, these DNA rotors produce sustained unidirectional rotations of over 20 revolutions per second. Speed control is available through setting the applied voltage or ionic concentration difference.

Our work presents a proof of principle for developing membrane-embedded artificial nanomachines that autonomously perform mechanical work, much like biological rotary motor proteins. Notably, the principle demonstrated here is not limited to the 6hb DNA structure, but any elastic beam with a sufficiently large charge should be able to spontaneously self-configure into a chiral shape that breaks symmetry needed for torque generation when docked onto a nanopore. In the current implementation, work done by the DNA rotor is used to move the rod through the viscous aqueous environment that is native to any nanoscale system, dissipating energy into the aqueous surroundings. In future applications, additional loads can be attached either to the top of the DNA rotor, or to a shaft that spans the membrane (the central dsDNA protrusion in the current design), so that the torque can be used for powering other processes. While this first generation of DNA rotary machines relies on the mechanical properties of the elastic DNA bundles and their spontaneous bending under local forces to create the torsional coupling mechanism that allows its drive, one may be able, with efforts, to expand this flow-driven mechanism to future designs of propeller-shaped rigid nanostructures with an intrinsically defined asymmetry. Such purposely designed chiral structures may be able to rotate and produce torque in predicted directions.

Like their biological equivalents, membrane-spanning nanoengines may in future be integrated into artificial or even biological systems and couple to chemical synthesis units like $F_1$ ATP Synthase to power mechanochemical catalytic reactions. Sea water versus freshwater salt gradients may be used to directly power nanoscale rotator motions. Benefitting from Micro-electromechanical systems (MEMS) techniques, it will be possible to fabricate large arrays of nanorotors that couple to one another and move in sync, to form large nanoscale energy-harvesting networks. These self-configured rotary nanomachines thus open up a range of novel directions in the field of nanoscale motors.

**Methods**

**Design, folding and purification of the DNA nanostructure**

The structure was designed using cadnano v0.2 using an 8064-bases long scaffold (sequences see Supplementary table S6) [26]. The folding reaction mixtures contained a final scaffold concentration of 50 nM and oligonucleotide strands (IDT) of 250 nM each. The folding reaction buffer contained 5 mM TRIS, 1 mM EDTA, 5 mM NaCl, and 10 mM $MgCl_2$. The folding solutions were thermally annealed using TETRAD (MJ Research, now Biorad) thermal cycling devices. The reactions were left at 65°C for 15 minutes and were subsequently subjected to a thermal annealing ramp from 60°C to 44°C (1°C/hour). The folded structures were stored at room temperature until further sample preparation steps. All folded structures were purified from excess oligonucleotides by PEG precipitation. All procedures were performed as previously described[27].

**Nanopore fabrication**

Nanopore arrays were fabricated by using Electron-Beam Lithography (EBL) and Reactive-Ion Etching (RIE)[28]. Before nanopore fabrication, 20-nm thick free-standing silicon nitride ($SiN_x$) membranes were fabricated as previously published[29]. Then, a layer of 100-nm thick polymethyl methacrylate (PMMA) electron sensitive resist (molecular weight 950k, 3% dissolved in anisole, MicroChem Corp) was spin-coated on top of the chip containing the freestanding $SiN_x$ membrane. Subsequently, the layer was patterned by exposing the resist with a 100 keV electron beam from the electron beam pattern generator (EBPG5200, Raith). The pattern was developed with a 1:3 mixture of methyl isobutyl ketone (MIBK) and isopropanol (IPA) for 1 min and stopped by immersing in IPA for 30 s. Finally, the exposed pattern was transferred to the $SiN_x$ membrane by RIE with fluoroform and argon (200 s, 50 W, 50 sccm of $CHF_3$, 25 sccm of Ar, 10 µbar, Sentech Plasma System SI 200), and the resist was stripped in oxygen plasma for 2min (200 $cm^3$/min $O_2$, 100 W, PVA Tepla 300) followed by an acetone bath for 5 min to remove the residual resist. Finally, the chips were rinsed with IPA and then spin dried.

**Fluorescent imaging of single rotors in nanopores**

Solid-state nanopore chips were cleaned in oxygen-plasma prior to every experiment (100 W, 50 mTorr, 1 min, Plasma Prep III, SPI Supplies). The nanopore chip was then mounted in a PDMS (SYLGARD™ 184 Silicone Elastomer) flow cell with a glass coverslip (VWR, No. 1.5) bottom (see Supplementary Figures S14). Coverslips were cleaned by sonication in acetone, IPA, deionized water (Milli-Q), 1 M KOH, deionized water for 30 min each, then drying thoroughly with compressed nitrogen and plasma cleaning under oxygen atmosphere. After assembly, the PDMS device was treated with 50 W oxygen-plasma for 3 min to increase its hydrophilicity before embedding a pair of Ag/AgCl electrodes, one in each reservoir, and flushing deionized water into the assembled flow cell to wet the system. The PDMS and the nanopore chips were always assembled shortly before the experiment and were single use only. The flow cell was then installed and fixed in a lab-built fluorescence microscope with a 60x water immersion objective (Olympus UPlanSApo 60x, NA1.20, water immersion) and a sCMOS camera (Prime BSI, Teledmy Photometrics). The camera field of view was reduced to the size required to allow data acquisition at high frame rates. The typical exposure time during each experiment was 2 ms with a corresponding frame rate of about 450-500 fps. The analysis is done using the framerate of 500 fps for simplicity.

For imaging, one end of DNA 6hb was labelled with 20x Cy5 tags. A 637-nm laser was used in epifluorescence mode. For two-color imaging, 20x Cy5 and 10x Cy3 tags were labelled on the two ends of each DNA structure. They were excited by 532-nm and 637-nm lasers simultaneously, and the two channels were split before the image acquisition and recorded on separate regions on the sCMOS camera. In this way, the two channels were imaged side-by-side and time-synchronized. Alignment of the two channels was done by a calibration procedure before or after each experiment, imaging a fiducial maker grid in the two channels and calculating the image transform function with a custom MATLAB script (see Data and Code Availability). The marker grid consisted of an array of subdiffraction-limited holes in a 100-nm Pd film with a 1.5 µm spacing filled with a mixture of Cy3 and Cy5 in solution. The typical alignment registration error was below 10 nm [30]. Prior to imaging the rotors, an imaging buffer (50 mM Tris-HCl pH 7.5; 50 mM NaCl, 5 mM $MgCl_2$; 1 mM DTT, 5% (w/v) D-dextrose, 2 mM Trolox, 40 µg ml$^{-1}$ glucose oxidase, 17 µg ml$^{-1}$ catalase; 0.05% TWEEN20) was flushed into the two reservoirs on both sides of the $SiN_x$ membrane. Electrodes were connected to a custom-built circuit to apply voltages[31]. DNA rotors were added into the electrically grounded compartment (*cis*) of the flow cell at a concentration of 1 pM.

Single DNA rotors were captured by nanopores in the array under a 100-mV bias voltage (unless otherwise stated) across the membrane. After video-data acquisition, the rotors were released from the nanopore array by reversing the bias voltage to -100 mV or lower and then setting it to 0 mV for several tens of seconds to allow the imaged rotors to diffuse away and avoid recapture of the same rotors in the next round of video recording. Then a new set of rotors was captured again by applying the positive bias voltage. To record the applied voltage for each optical frame in the voltage-response experiments, the camera's logic output signal (voltage 'on' when the camera is exposing the first row of pixels of a frame) and the measured applied voltage signal from the lab-built circuit were simultaneously sampled with a DAC board (USB-6251, National Instruments) and transferred to a computer, processed, displayed and stored with an in-house LabVIEW control program.

In transmembrane salt gradient experiments, the *cis* reservoir contained imaging buffer with a lower salt concentration (50 mM NaCl), and the *trans* reservoir with a higher salt concentration (typically 550 mM NaCl). The data acquisition and imaging process was identical to the voltage-driven experiments.

**Single-particle localization and rotor data analysis**

Acquired image sequences were first processed by Fiji (ImageJ) with the ThunderStorm plugin[32] for single-molecule localization. The diffraction-limited point spread function was filtered after applying a wavelet filter (B-spline) to the images. Then approximate localization of molecules was determined by using the local intensity maximum and the sub-pixel position of the fluorescent molecules was subsequently determined using an integrated gaussian method. Post-localization processing was done with a quality filter (localization uncertainty <50 nm) to remove spuriously detected spots, drift-correction (using transcorrelation method) was applied, and a density filter (requiring at least 15 detected spots in a 50nm-in-diameter radius around each detected spot in the entire video sequence) was used to validate the detected spots.

The single-molecule localization results were then analyzed using in-house MATLAB scripts (see Code Availability). First, we clustered all the coordinates from ThunderStorm into spatially local groups based on their Euclidean separation using MATLAB's Computer Vision toolbox. Each group contained the positions of the tip of a single rotor.

Next, we fitted a circle to the data for each rotor, the center of which was used as a reference point to calculate the angular displacement between frames. We then inspected each data group separately and only considered the rotors that traced at least part of the fitted circle for the next stage of analysis. Usually, about 30% of rotor data groups were retained for further analysis. The remainder was excluded as they displayed no motion either due to nonspecific binding between DNA and the surface, multiple rotors inserting into the same pore (as observed by a stepwise intensity increase), or low data quality. The steps were determined by calculating the angles between the fluorophore positions in adjacent frames. If the angle was smaller than or equal to 180° it was counted as a counterclockwise step, and if the angle was larger than 180° it was counted as a clockwise step. Finally, we calculated the cumulative angular displacements and the mean-square displacements (MSD) for each rotor to analyze their motion. The velocity $v$ was determined by fitting $MSD = v^2\tau^2 + 2D\tau$ to the MSD curve of each rotor.

**Atomic Force Microscopy imaging**

A stock solution of DNA 6hb structure was first diluted in Milli-Q water, supplemented with 7.5 mM of MgCl$_2$, to a final concentration of 25 pM. The mixture with the DNA rotor structure was deposited onto freshly cleaved mica and incubated on top of it for 30 seconds. Then, the surface was thoroughly washed with 3 ml of Milli-Q water and dried under an air-flow[33]. Images were taken with a multimode AFM from Bruker (Bruker corporation, Massachusetts, USA) using ScanAsyst-Air tips from Bruker. The AFM was operated using peak-force QNM mode for imaging in air, at room conditions. Image processing and data extraction was done with WSxM software[34].

**Discrete elastic rod model**

For the simulations, we set a fixed length of $L = 450$nm for the 6hbs whose shape was approximated by a set of $N = 51$ vertices in three dimensions. For the bending stiffness, a persistence length of $L_p = 1.5$μm was adopted. For the various interactions, we computed the bare charge density of the 6hb as $q_{\text{bare}} = \frac{6(2e^-)}{a} = 5.7 \cdot 10^{-9} \frac{\text{C}}{\text{m}}$, with α=0.34 nm, and we introduced a control parameter $0 \leq \lambda_{\text{eff}} \leq 1$ for the effective electrophoretic mobility. The 6hbs experience a drag force when moving through the fluid which is captured by the friction coefficients ζ$_\perp$ and ζ$_\parallel$, with $\frac{\zeta_\perp}{l} = \frac{2\zeta_\parallel}{l} = \pi\eta$ per rod length $l = 30$nm, with the viscosity of water $\eta = 1$mPas. To estimate the relative strength of the electrophoretic and electroosmotic effects, we employed two control parameters, $\lambda_{\text{eff}}$ (see above) and a flow factor $\lambda_{\text{flow}}$ that characterizes the strength of the osmotic flow. For the examples in Fig. 3, $\lambda_{\text{eff}} = 0.26$ and $\lambda_{\text{flow}} = 19.5$ were used. As both electroosmosis and electrophoresis scale with the external electric field, it is most relevant to consider the ratio of $\lambda_{\text{flow}}$ and $\lambda_{\text{eff}}$. A scan over various parameter values is displayed in Fig. S11. The data showed a robust parameter regime with stably rotating configurations. Examples for various initial conditions under a single set of parameters are shown in Supplementary Video 2-6.

**Acknowledgements**

We thank Aleksei Aksimentiev, Christopher Maffeo, Miloš Tišma, Alessio Fragasso and Anders Barth for discussions, Biswajit Pradhan for help with single-molecule fluorescence setup, Nils Klughammer for help with the fabrication of fiducial marker grids for dual-channel fluorescence imaging, and Philip Ketterer for initial DNA origami structure designs. We acknowledge funding support by Dutch Research Council NWO grant NWO-I680 and the European Research Council Advanced Grant 883684 (CD). This work was supported by a European Research Council Consolidator Grant to H.D. (GA no. 724261), the Deutsche Forschungsgemeinschaft through grants provided within the Gottfried-Wilhelm-Leibniz Program (to H.D.), and the SFB863 Project ID 111166240 TPA9 (to H.D.). The work has received support from the Max Planck School Matter to Life (to R.G. and H.D.) and the MaxSynBio Consortium (to R.G.), which are jointly funded by the Federal Ministry of Education and Research (BMBF) of Germany, and the Max Planck Society.

**Author contributions.**

X.S., D.V., H.D. and C.D. conceived the concept of DNA rotors in nanopores. A-K.P. and H.D designed and prepared the DNA origami structures. X.S designed the nanopore experiment and fabricated nanopore devices. X.S. and W.Z. conducted nanopore experiments. A.M.G. performed AFM measurements. X.S and D.V. wrote the data analysis program and analysed data. J.I. and R.G. designed and conducted theoretical modelling and simulations. All authors discussed experimental findings and co-wrote the manuscript.

**Competing interests**

The authors declare no competing interests

**Data availability**

All experimental data are available at http://doi.org/10.5281/zenodo.6513594.

**Code availability**

MATLAB codes for data processing are available at:

    http://doi.org/10.5281/zenodo.6513594

Julia codes used for numerical simulation are available at:

    https://gitlab.gwdg.de/LMP-pub/nanoturbines

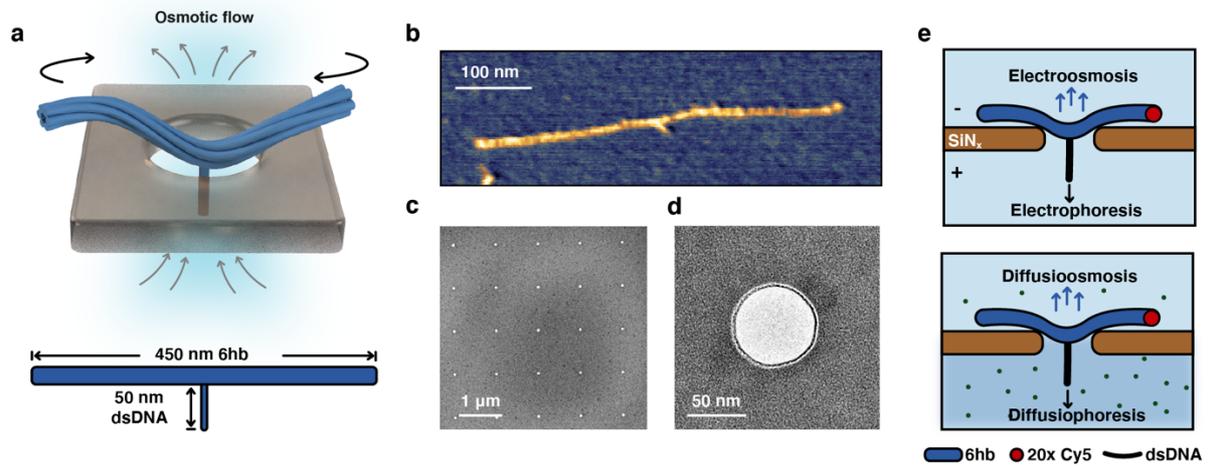

**Fig. 1. DNA rotor on a solid-state nanopore. (a)** Schematic of a DNA rotor on a nanopore. A 450nm long 6-helix DNA bundle is docked onto a ~50 nm diameter nanopore, where an electro- or diffusiophoretic force locally bends and deforms the bundle, whereupon an osmotically induced water flow causes it to rotate. **(b)** AFM image of a DNA rotor structure. **(c)** TEM image of a nanopore array. **(d)** TEM image of a single nanopore. **e,** Schematics of the two ways to drive the rotor. Top: Voltage-driven drive, where the DNA is driven towards the nanopore via electrophoresis and couples to the oppositely directed electroosmosis flow from the nanopore. Bottom: ion-gradient driven, where, analogously, the DNA is driven towards the nanopore by diffusiophoresis and powered by the diffusioosmotic flow running in the opposite direction.

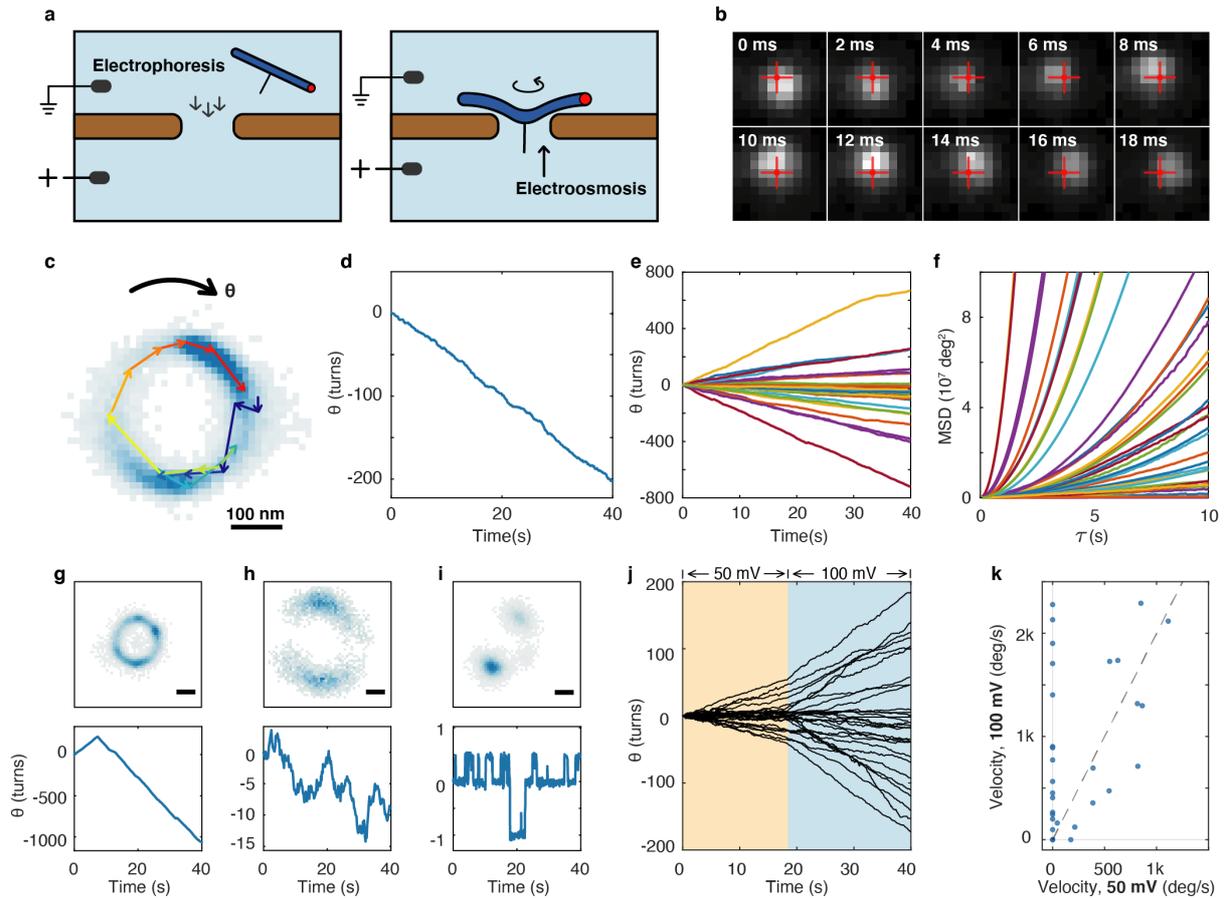

**Fig. 2. Unidirectional rotation of a DNA rotor under a transmembrane voltage.**
**(a)** Schematic of a DNA rotor docking onto a nanopore. **(b)** Example of consecutive fluorescence images (2 ms exposure time) of one tip of a single rotor that is fluorescently labelled, demonstrating rotation about its center point in the nanopore which is indicated with the red cross. **(c)** Corresponding heatmap (blue pixels) of single-particle localizations for the rotor in panel b. An example trajectory of 20 subsequent positions of the labelled tip is overlaid. **(d)** Corresponding cumulative angle versus time for the same rotor, showing a rotation over hundreds of turns. **(e)** Example cumulative angular-displacement curves for 40 rotors. **(f)** Corresponding MSD curves. **(g-i)** Examples of localization heatmaps and corresponding cumulative angular displacement curve for rotors with different behaviors than displayed in panel c-d, viz., a rotor that changed direction (g), a rotor dominated by Brownian motion (h), and a rotor that toggles between two diametrically opposite positions (i). All scale bars 100 nm. **(j)** Cumulative angular displacement for 30 rotors, where the applied transmembrane voltage was changed from 50 mV to 100 mV after 18 seconds. **(k)** Corresponding average rotation velocities of the rotors in panel j, before (50 mV) and after (100 mV) the applied voltage was increased. The dashed line denotes $V_{100mV} = 2 \times V_{50mV}$.

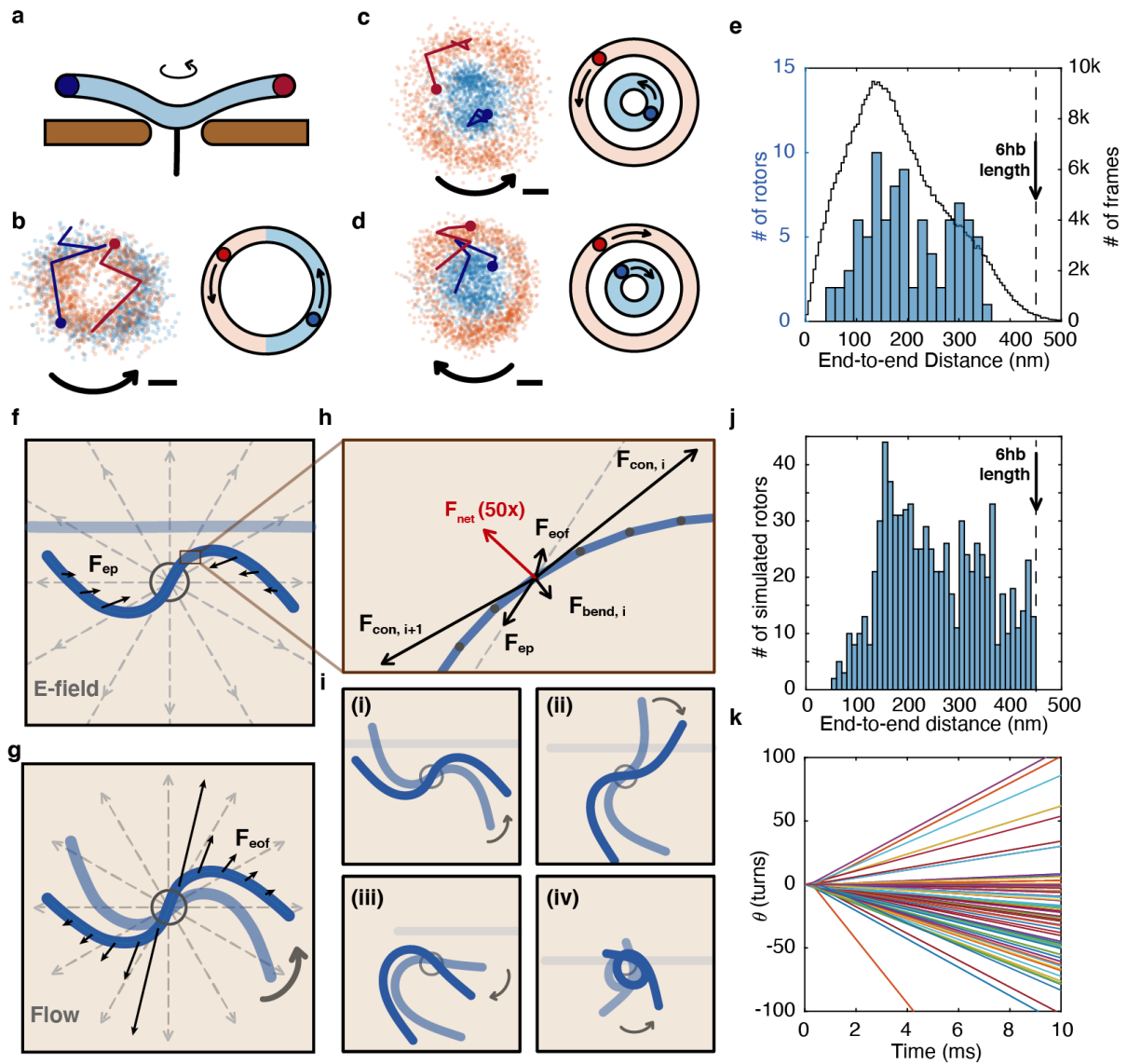

**Fig. 3. DNA rotors are deformed by the E field and driven to rotation by the flow. (a)** Side view schematic of a dual-color labelled DNA rotor docking onto a nanopore. **(b-d),** Scatter plots of localization coordinates for the two rotor tips that are labelled in different colors (left) and schematics of the corresponding moving patterns (right). The diameters of the circular trajectories are observed to be either the same or different from both ends. Both symmetric and asymmetric trajectories are observed for the two tips, which are attributed to on- or off-center docking of the rotors, respectively. Scale bars are 100 nm. **(e)** Histogram of the average end-to-end distance during rotation of all analyzed rotors (blue, n=90) and histogram of end-to-end distance measured in each frame (black, n=425486). The majority of 6hbs exhibit an end-to-end distance that is much shorter than the 450-nm designed length, revealing that the rotors are mechanically deformed upon docking. **(f)** Schematic that illustrates how a DNA 6-helix bundle deforms under a compressive electrophoretic force. **(g)** Schematic that illustrates how coupling of the radial electroosmotic flow to non-radially aligned rod segments produce a sustained rotation. As the perpendicular friction coefficient is about twice larger than the tangential coefficient, the resulting force on the bundle drives the structure in the angular direction. **(h)** Force balance diagram (to scale) for the rod segment in the rectangular box in panel e, showing that internal and external forces add up to a non-vanishing force that drives a rotation. $F_{con}$: the internal constraint forces; $F_{bend}$: the induced mechanical bending force; $F_{ep}$: the effective electrophoretic force; $F_{eof}$: drag force generated by electroosmotic flow. For visual clarity, the net resulting force (red arrow, rotation-induced drag force not included) is drawn 50 times longer as compared to the other force vectors. Black dots denote the ends of each rod segment. **(i)** Examples of the result of simulations where bent beams produced a sustained rotation, displayed in top (XY) view. Corresponding XZ views are shown in Fig. S7. Light, medium, and darker blue shades denote a time sequence of the initial placement to bending and rotation. **(j)** Histogram of end-to-end distance during rotation extracted from the simulations of all rotating simulated rods in the phase-space diagram (n=810, Fig. S8) **(k)** Cumulative angular displacement curves for 417 DNA bundles, obtained from simulations.

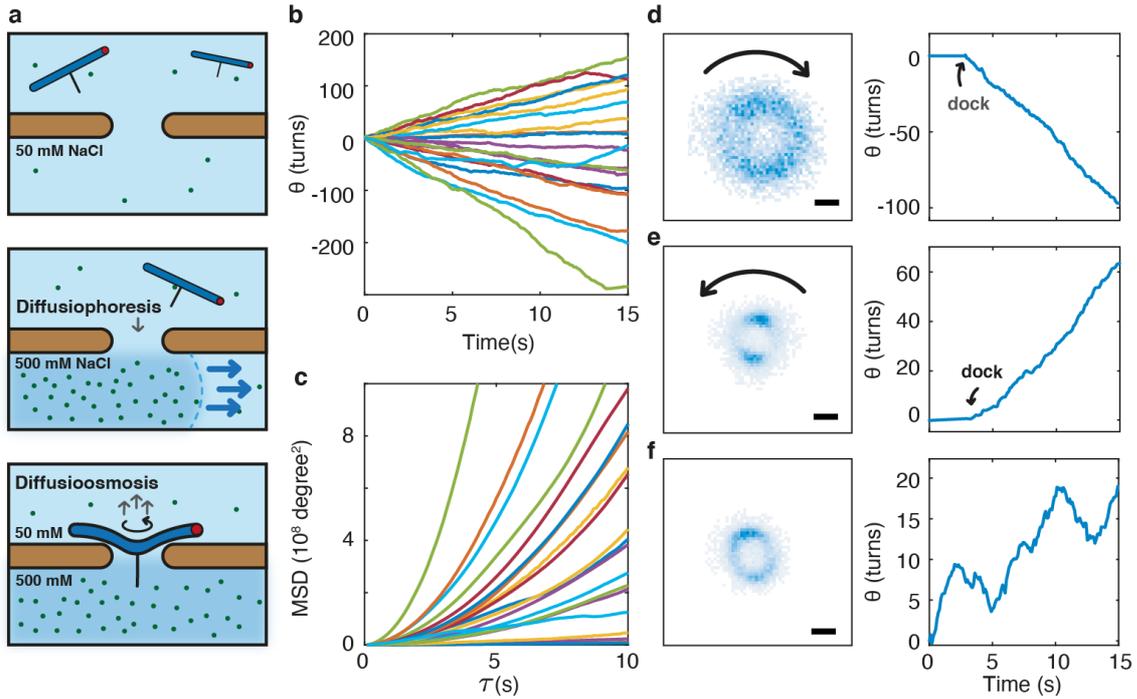

**Fig. 4. Unidirectional rotation of rotors driven by a transmembrane salt gradient.** **(a)** Schematic of a DNA rotor that is docking onto a nanopore and driven by an osmotic flow induced from a static transmembrane salt gradient with 50mM NaCl in the cis reservoir and 550mM NaCl in the trans reservoir. **(b)** Cumulative angular displacement curves of 20 different rotors. **(c)** Corresponding MSD curves. **(d-f)** Examples of localization heatmaps (left) and corresponding cumulative angular-displacement curves (right) for rotors driven by a transmembrane salt gradient. Where panels (d) and (e) show examples of a nanorotors demonstrating driven rotation, (f) shows an example of a rotor for which Brownian motion dominates. All scale bars are 100 nm.

# Supplementary Materials for

## Sustained unidirectional rotation of a self-organized DNA rotor on a nanopore

Xin Shi,[1] Anna-Katharina Pumm,[2] Jonas Isensee,[3] Wenxuan Zhao,[1] Daniel Verschueren,[1, †] Alejandro Martin-Gonzalez,[1] Ramin Golestanian,[3,4, *] Hendrik Dietz,[2,*] Cees Dekker[1,*]

[1] Department of Bionanoscience, Kavli Institute of Nanoscience Delft, Delft University of Technology, Delft, The Netherlands.
[2] Department of Physics, Technical University of Munich, Garching 85748, Germany
[3] Max Planck Institute for Dynamics and Self-Organization, Göttingen 37077, Germany
[4] Rudolf Peierls Centre for Theoretical Physics, University of Oxford, OX1 3PU, Oxford, UK
† Current address: The SW7 Group, 86/87 Campden Street, W8 7EN, London, UK
* Corresponding authors. Emails: c.dekker@tudelft.nl, dietz@tum.de, ramin.golestanian@ds.mpg.de

**This PDF file includes:**

> Supplementary Notes 1 to 4
> Figs. S1 to S13
> Tables S1 to S5
> Captions for Movies S1 to S6

**Other Supplementary Materials for this manuscript include the following:**

> Movies S1 to S6
> Table S6

# Supplementary Notes

## 1. COMSOL simulation of the electroosmotic flow

We estimate the electric field electroosmotic flow field distribution using a simplified model solved by COMSOL Multiphysics. The ion flux under the influence of both an ionic concentration gradient $\nabla c$ and an electric field was described by the Poisson-Nernst-Planck equation and electroosmotic flow was described via the Navier-Stokes equation for incompressible fluid without inertial terms [1] because of the small Reynolds number ($\sim 10^{-3}$-$10^{-4}$) in our nanoscale geometry.

Figure S12 shows the geometry of the model. Table. S1-5 shows the key boundary conditions used in simulation. Fig. S13 shows the simulation results.

## 2. Discrete elastic rod model for the field-induced shape deformation and flow coupling of the origami rotors

The overdamped dynamics of the rod and its velocity $v$ are described by a force balance, as illustrated in Fig. 3G and in the form of the equation

$$0 = Zv + F_{\text{bend}} + F_{\text{ep}} + F_{\text{eof}} + F_{\text{con}},$$

with

$$Z = \begin{pmatrix} \zeta_{\parallel} & & \\ & \zeta_{\perp} & \\ & & \zeta_{\perp} \end{pmatrix},$$

which holds for all points along the rod. Here $\zeta_{\parallel}$ and $\zeta_{\perp}$ are the friction coefficients along the principal axes of the rod, along and normal to its tangent vector $\widehat{T}$. We assume these to be given by resistive force theory as $\zeta_{\perp} = 2\zeta_{\parallel} = \pi \eta l$ with the viscosity of water $\eta$ and rod segment length $l$. The force terms are the force due to the rod's bending resistance ($F_{\text{bend}}$), the electrophoretic force due to the external electric field ($F_{\text{ep}}$) acting on the charged DNA origami, the drag force due to the external electroosmotic flow ($F_{\text{eof}}$), and the strain force between the rod's elements ($F_{\text{con}}$).

To simulate the dynamics of the DNA origami rod, we make use of the established discrete elastic rod (DER) description *(ref 22 in the main text)*. We treat the bundle as an isotropic rod with a straight resting configuration and assume inextensibility. We then discretize the rod at equally spaced vertices $x_1$ to $x_N$, yielding $N - 1$ displacement vectors $e_i = x_{i+1} - x_i$.

Forces resulting from the external electric field are the electrophoretic contribution

$$F_{ep} = \lambda_{\text{eff}} q_{\text{bare}} E,$$

where $q_{\text{bare}}$ is the bare charge per segment of the 6 hb (a line charge density of $\rho_{\text{bare}} = \frac{6(2e^-)}{a}$, with $e$ the elementary charge and $a$ the distance (0.34 nm) between two consecutive bases), $E$ the external electric field at the segment's location and $\lambda_{\text{eff}}$ is a dimensionless scale factor between 0 and 1 that measures the strength of the electrophoretic mobility. This scale factor describes the effect of counterion condensation, known to be present when divalent counterions ($Mg^{2+}$) are present in the electrolyte solution [2]. Local electroosmotic flows induced by the DNA are accounted for by contributions that purely oppose the electrophoretic pull *(ref 18 in the main text)* (included in $\lambda_{\text{eff}}$) and ones that enhance the background electroosmotic flow (represented by $\lambda_{\text{flow}}$). Typical values for $\lambda_{\text{eff}}$ range from 0.05 to 0.25 and $\lambda_{\text{flow}}$ between 10 and 25).

The electro-osmotic drag force that the radial flow with flow velocity $\boldsymbol{v}_{\text{eo}}$ from the pore exerts on the rod segment is described as

$$\boldsymbol{F}_{\text{eof}} = \boldsymbol{Z}\boldsymbol{v}_{\text{eo}},$$

with $\boldsymbol{Z}$ as described above. Both the electric field $\boldsymbol{E}$ and the background flow $\boldsymbol{v}_{\text{eo}}$ are inputs into the model and determined beforehand using finite-element COMSOL simulations (Supplementary Note 1) of the nanopore system in the absence of the DNA rod.

The bending force on the rod segment can be derived from the bending energy $U_{\text{bend}}$ that is defined as an integral of the curvature over the length of the rod $L$ with persistence length $L_p$,

$$U_{\text{bend}} = \frac{L_p k_B T}{2} \int_0^L dl\, \kappa^2,$$

with $k_B$ the Boltzmann constant, $T$ temperature, and $\kappa$ curvature. The best approximation of the resulting forces on discrete vertices was derived in *(reference 22 in the main text)*. In the special case of a torsion-free rod, this becomes

$$F_{\text{bend},i} = \frac{-2L_p k_B T}{l} \sum_{j=i-1}^{i+1} \boldsymbol{\nabla}_i (\kappa\boldsymbol{b})_j^T (\kappa\boldsymbol{b})_j,$$

with bending stiffness $L_p k_B T$, the segment length $l = \frac{L}{N-1}$ (contour length $L$ of the total rod, $N$ the number of vertices in the rod), and $\kappa\boldsymbol{b}$ the curvature multiplied by the (normalized) binormal vector

$$(\kappa\boldsymbol{b})_i = \frac{2\boldsymbol{e}^{i-1} \times \boldsymbol{e}^i}{l^2 + \boldsymbol{e}^{i-1} \cdot \boldsymbol{e}^i}.$$

The gradients are given by

$$\boldsymbol{\nabla}_{i-1}(\kappa\boldsymbol{b})_i = \frac{2[\boldsymbol{e}^i] + (\kappa\boldsymbol{b})_i (\boldsymbol{e}^i)^T}{l^2 + \boldsymbol{e}^{i-1} \cdot \boldsymbol{e}^i}$$

$$\boldsymbol{\nabla}_{i+1}(\kappa\boldsymbol{b})_i = \frac{2[\boldsymbol{e}^{i-1}] + (\kappa\boldsymbol{b})_i (\boldsymbol{e}^{i-1})^T}{l^2 + \boldsymbol{e}^{i-1} \cdot \boldsymbol{e}^i}$$

$$\boldsymbol{\nabla}_i(\kappa\boldsymbol{b})_i = -(\boldsymbol{\nabla}_{i-1} + \boldsymbol{\nabla}_{i+1})(\kappa\boldsymbol{b})_i$$

where $[\boldsymbol{a}]$ is a skew-symmetric 3×3 matrix such that $[\boldsymbol{a}] \cdot \boldsymbol{x} = \boldsymbol{a} \times \boldsymbol{x}$.

In the approximation of individual connected straight rods, the tangent vector becomes ill-defined at the vertices. Therefore, we compute $\boldsymbol{F}_{\text{ep}}$ and $\boldsymbol{F}_{\text{eof}}$ at the midpoints of every segment and distribute the result evenly between the two enclosing vertices. To approximate the mobility of each vertex we approximate the tangent by the secant connecting the neighbouring vertices $\widetilde{\boldsymbol{T}}_i = \frac{x_{i+1} - x_{i-1}}{|x_{i+1} - x_{i-1}|}$ assuming a vanishing curvature at both ends.

One further addition to the model is needed to stabilize the dynamics, which is to account for the finite thickness of the rod. We added a steep (one-sided) quadratic potential along the vertical direction to keep the rod from penetrating the membrane. Time integration is done using an adaptive Euler scheme ensuring that no vertex moves by more than $\approx 20$ pm per step.

Central results of these simulations are shown in Fig. 3 with a phase diagram in Extended Data Fig. 8. Important features of the experiment such as docking, and translocation are reproduced in the case of strong electric fields. Additionally, there exists a parameter regime that exhibits stably rotating configurations that reproduce many of the experimentally observed end-to-end distances as well as asymmetrically distributed endpoints. Examples for various initial conditions under a single set of parameters are shown in Supplementary Video 2-6.

## 3. Estimation of the max salt diffusion rate across the nanopore array with 550mM:50mM salt gradient

The maximum rate of concentration change in the flow cell during the ion-gradient driven experiments can be estimated with Fick's law

$$J = -D \frac{\partial c}{\partial x},$$

where $D$ is the diffusion coefficient of the ions, and c is the concentration gradient, and x the coordinate along the pore axis. In the simplified scenario of salt gradient across a nanopore, we can simplify the equation as

$$J = -D \frac{\Delta c}{L},$$

where $\Delta c$ is the concentration difference on the two sides of the membrane, $L$ is the thickness of the membrane. Using $D = 1.6 \times 10^{-9}$ m²/s, $L = 20 \times 10^{-9}$ m, $\Delta c = 450$ mM we obtain a flux $J \sim -25$ mol/m²s. With the diameter of the nanopore $d = 50$ nm, the volume of the two reservoirs at 50 μL, and 400 nanopores on a typical membrane, we can estimate that the maximum concentration change rate is 0.6 μM/s. This is equivalent to a concentration change of about 1mM concentration change per 28 min, causing only an insignificant change of less than 1% in the concentration gradient in a timescale way longer than our typical experiment.

## 4. Estimation energy consumption rate of rotors

We estimate the rotational friction coefficient for the rotor as

$$\xi_r = 3\pi \int_0^L dl r^2 \left[ \xi_\parallel (\hat{T} \cdot \hat{e}_\varphi)^2 + \xi_\perp \left(1 - (\hat{T} \cdot \hat{e}_\varphi)^2 \right) \right]$$

where $r$ denotes the distance to the center of rotation, $\hat{T}$ the tangent vector along the rod with $\xi_\perp = 2\xi_\parallel = \eta = 1$ mP s , and $L = 450$ nm. Using stationary conformations from the simulations we compute $\xi_r \approx 0.012$ pN nm s. Based on the maximum observed angular velocity of 125 radians/s we can estimate an energy consumption rate of 190 pN nm/s (~45 $k_B$T/s).

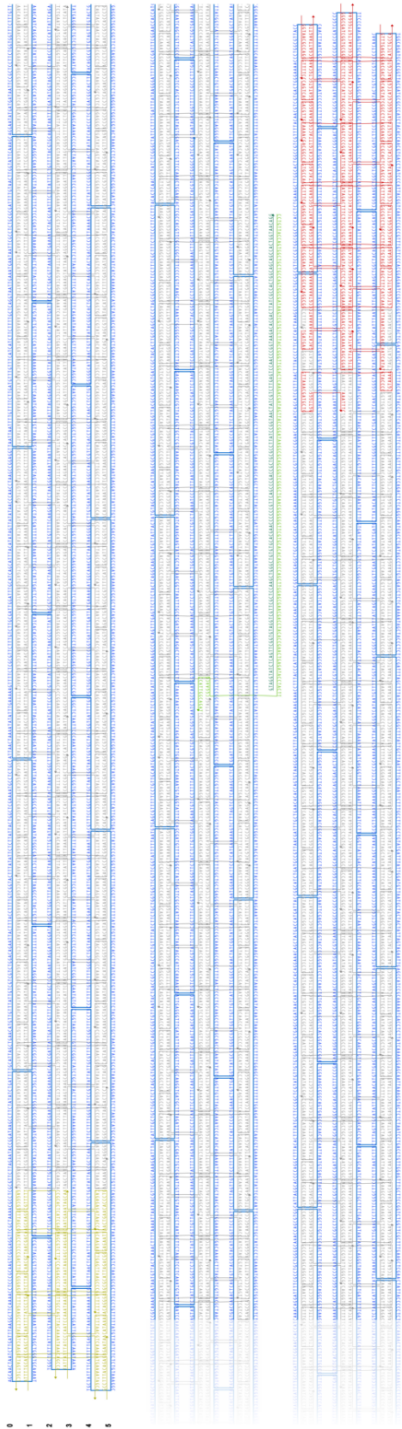

**Fig. S1. Diagram of the DNA rotor structure**

caDNAno (v0.222) scheme of the DNA structure. Fluorescent dyes are attached to the 5' ends of the colored strands (yellow: 10x Cy3 dyes span over ~20nm, red: 20x Cy5 dyes span over ~40 nm).

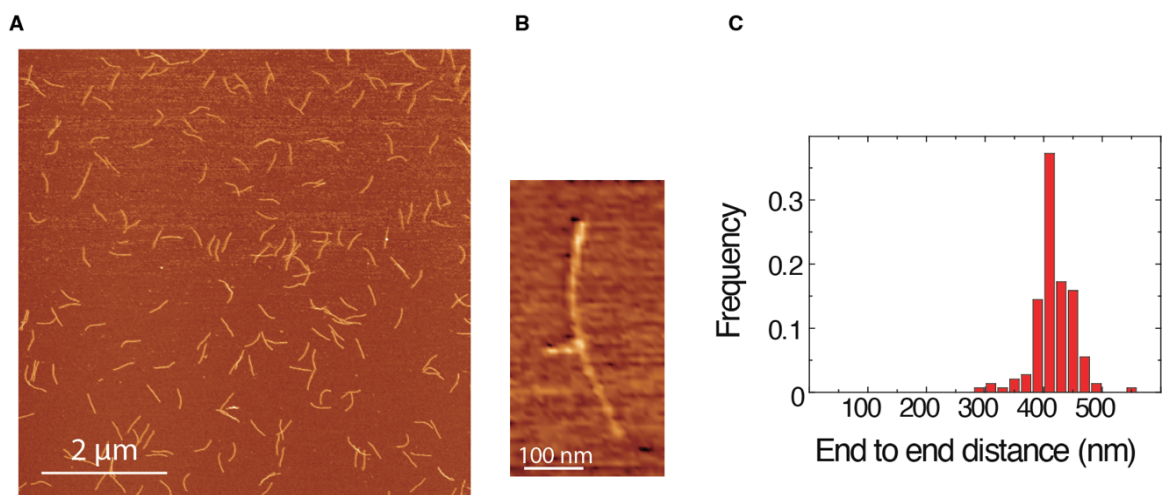

**Fig. S2. Atomic Force Microscopy (AFM) images of the DNA structures.**

**(A)** AFM image of the structures. **(B)** Zoom-in AFM image of an example of the structure. **(C)** Histogram of the end-to-end distance of the structures measured from AFM images. AFM images were calibrated using the known full contour length of the DNA 6hb (450 nm). AFM images show the six-helix bundle structure in a linear shape with, in some cases, a small protrusion from the middle. The relatively low number of origami where the 50 nm dsDNA leash protrusion is resolved can be explained by the random deposition of the structure on the surface, which sometimes hides or masks the protruding leash.

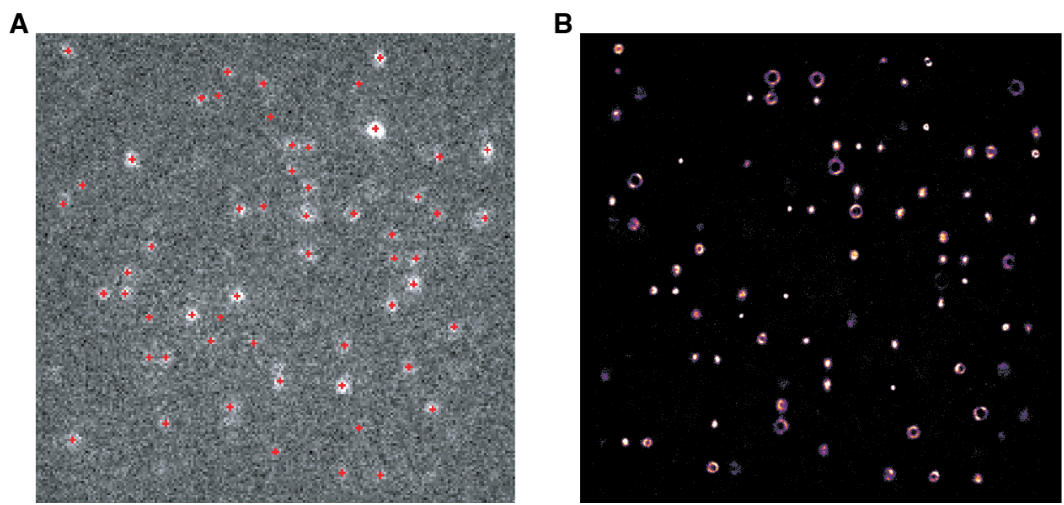

**Fig. S3. Example of the high-throughput fluorescence imaging of rotors in a nanopore array.**

**(A)** Example raw data frame of DNA rotors docked onto a nanopore array, and the corresponding single-particle localization results. Frame exposure time 2 ms. **(B)** Results of the single-particle localization from 8000 frames. Both stationary (points) and rotary (circles) rotors are observed.

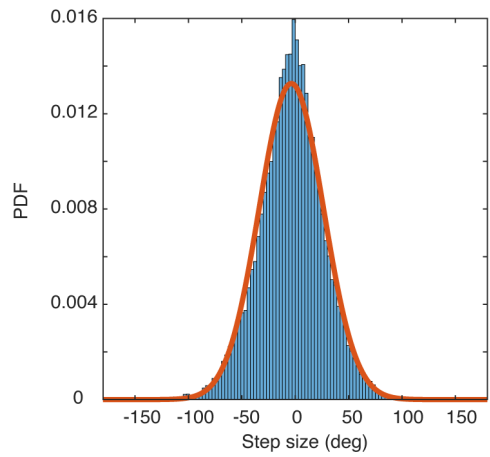

**Fig. S4. Typical distribution of the step sizes (corresponding to Fig 2d in the main text).** A normal distribution was fitted to the data which produced a mean of -3.6° and a standard deviation of 30.1°.

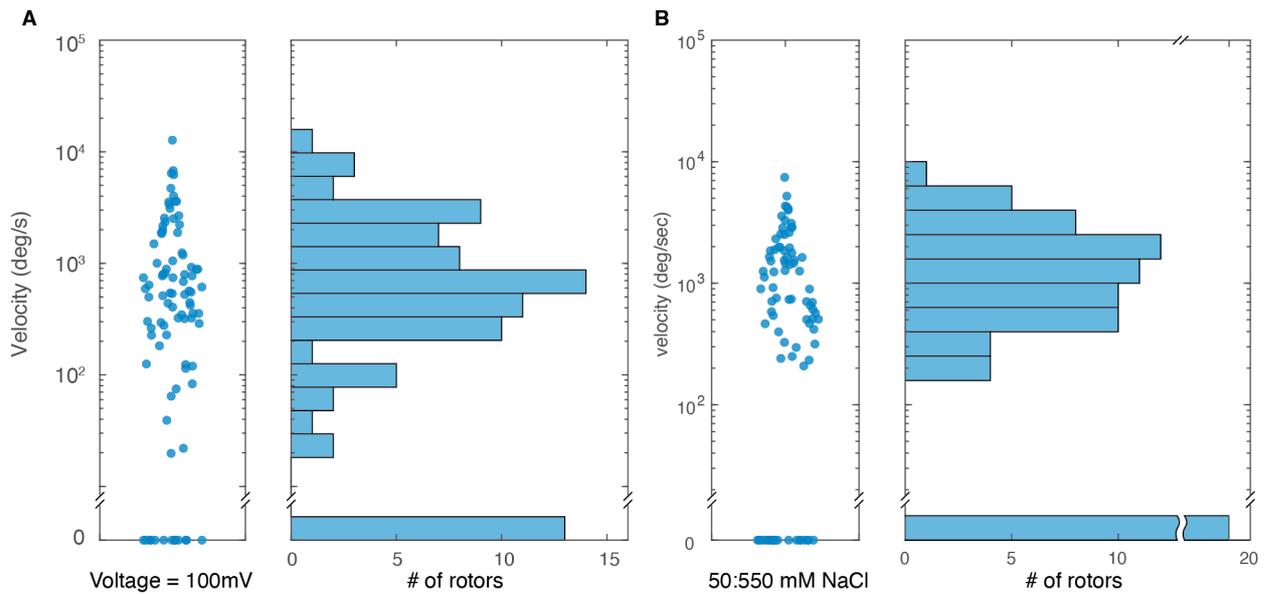

**Fig. S5. Rotor velocities from a typical voltage-driven experiment and a typical ion-gradient-driven experiment.**

**(A)** Histogram of rotation velocities under 100mV bias voltage. The velocity $v$ was determined by fitting $MSD = v^2\tau^2 + 2D\tau$ to the MSD curve of each rotor (as shown in Fig. 2F). Among a total of 600 docked structures in the nanopore array (from 3 batches of docking), 89 of them were included in the final analysis. Among these 89 rotors, 37 showed a velocity higher than 720 deg/s, 33 showed a velocity between 100-720 deg/s, and 13 showed no driven motion (indicated as $v = 0$ here) but instead a constrained Brownian motion. **(B)** Histogram of rotation velocities for experiments with a 50:550mM NaCl transmembrane salinity gradient. The velocity $v$ was determined by fitting $MSD = v^2\tau^2 + 2D\tau$ to the MSD curve of each rotor (as shown in Fig. 4C). Among a total of 468 docked structures in the nanopore array (from 2 batches of docking), 84 of them were included in the final analysis. Among these 84 rotors, 43 showed a velocity higher than 720 deg/s, 22 showed a velocity between 100-720 deg/s, and 19 show no driven motion (indicated as $v = 0$ here) but constrained Brownian motion instead.

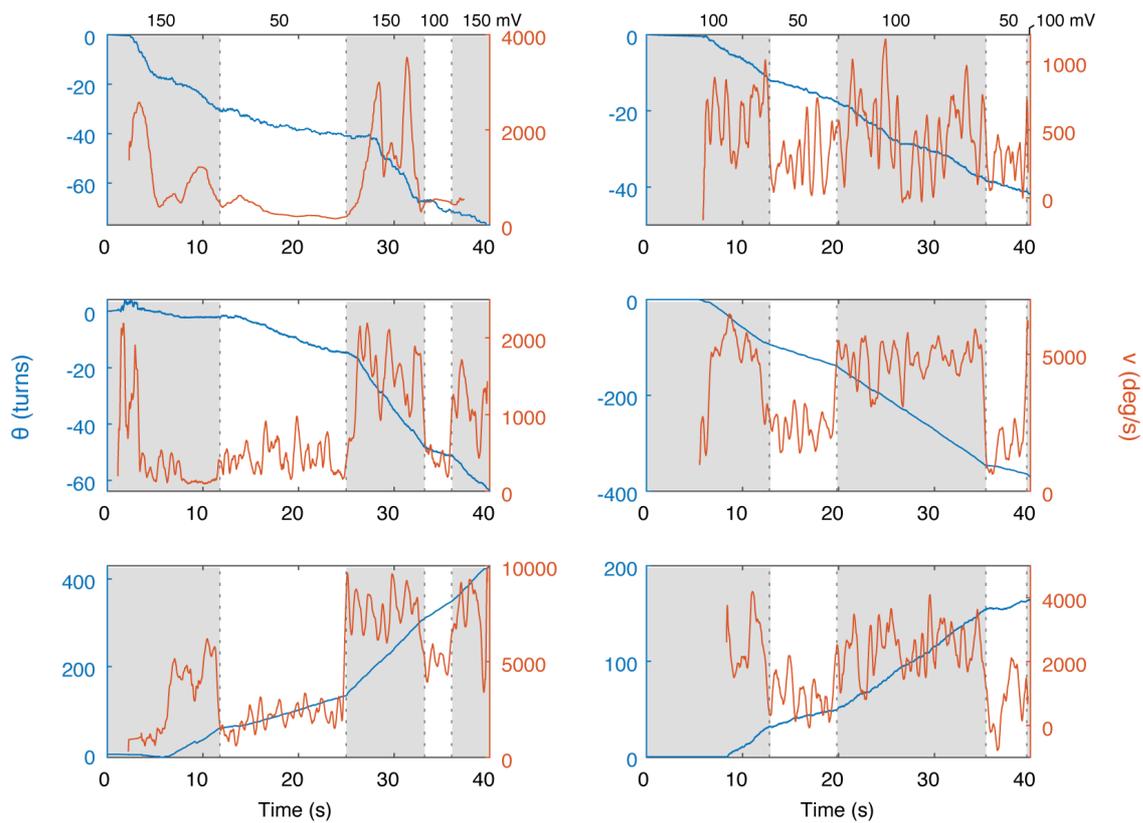

**Fig. S6. Examples voltage-driven rotor behaviours with an applied bias voltage that changed over time.**

Data in blue (red) show the rotation angle (speed) versus time for applied voltages as denoted black on the top. The rotational speed typically reverts to a value similar to the original value after the high voltage was re-established.

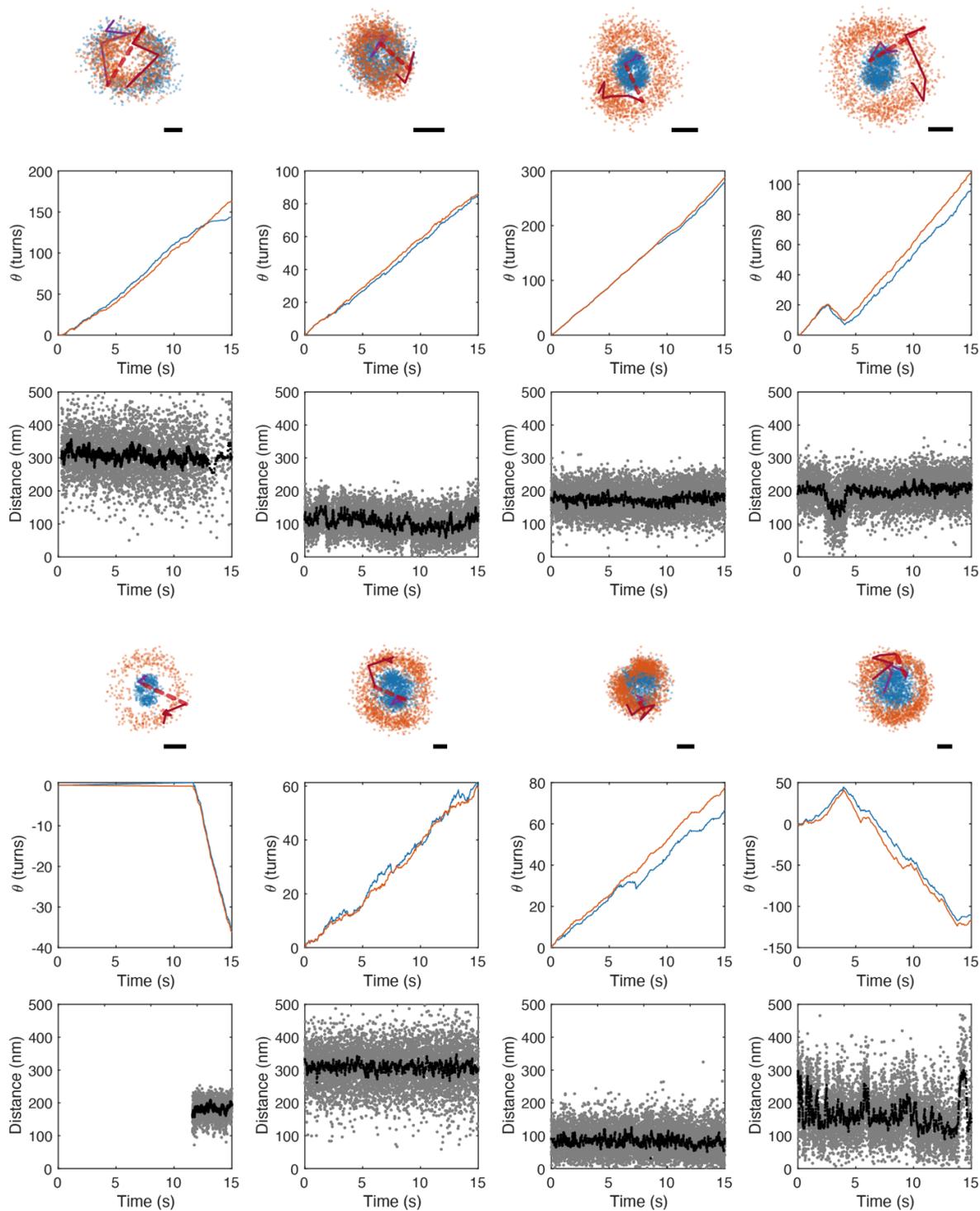

**Fig. S7. Examples of 2-color-labelled voltage-driven rotors.**

Each example includes at the top: a scatter plot of localization coordinates for two-color-labelled rotors (i.e., with Cy5 and Cy3 on each end respectively); at the middle: a corresponding cumulative angular displacement curve of both ends to show the driven rotation; and at the bottom: the corresponding end-to-end distance over time (grey) with a smoothed trace using a moving median filter with a window size of 40 data points (black).

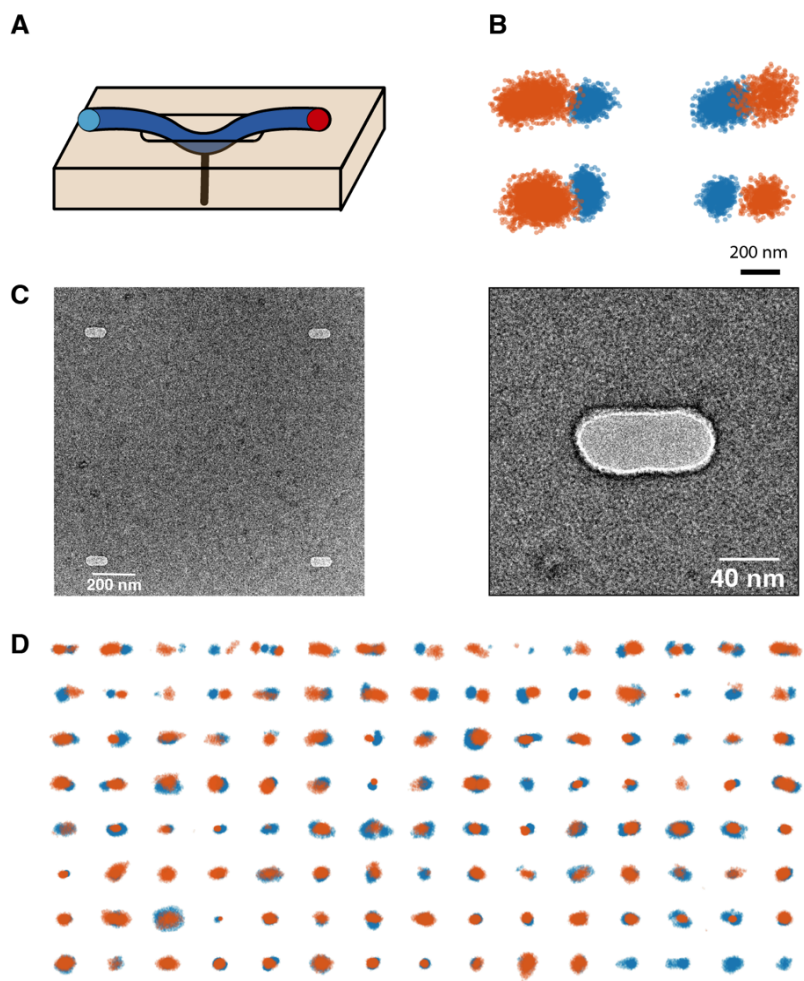

**Fig. S8. Docking DNA rotors on elongated nanopores.**

**(A)** Schematic of the docking experiment. Similar to the experiment shown in Fig. 3, the two ends of DNA T-rotors were labelled with 10x Cy3 and Cy5 fluorophores at the two ends respectively, and the rotors were docked onto the nanopore array by applying 100 mV bias voltage across the $SiN_x$ membrane. Typically, the motion of these nano rotors was stalled and the two-ends aligned with the long-axis of the nanopore. **(B)** Examples of the single-particle-localization results of the docked structures on elongated nanopores. **(C)** TEM image of the elongated nanopore array (left) and a zoom-in view of a single elongated nanopore (right). The length of this intentionally noncircular nanopore is about 80 nm while the width is about 40 nm. **(D)** More examples of the docked DNA rotors on elongated nanopores.

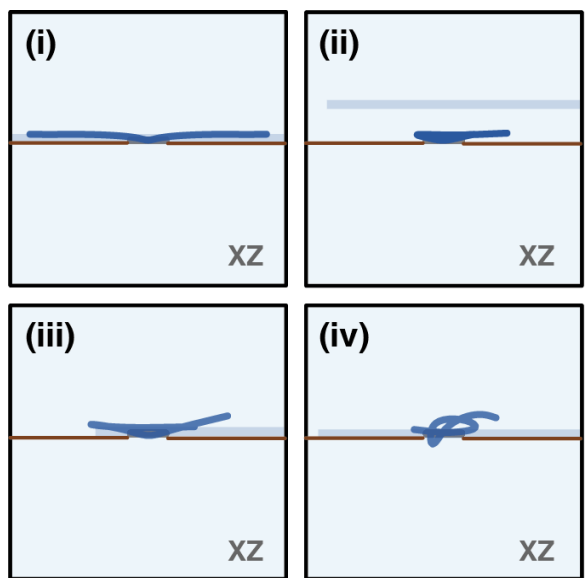

**Fig. S9. Corresponding XZ views of simulated rods in Fig 3i.**

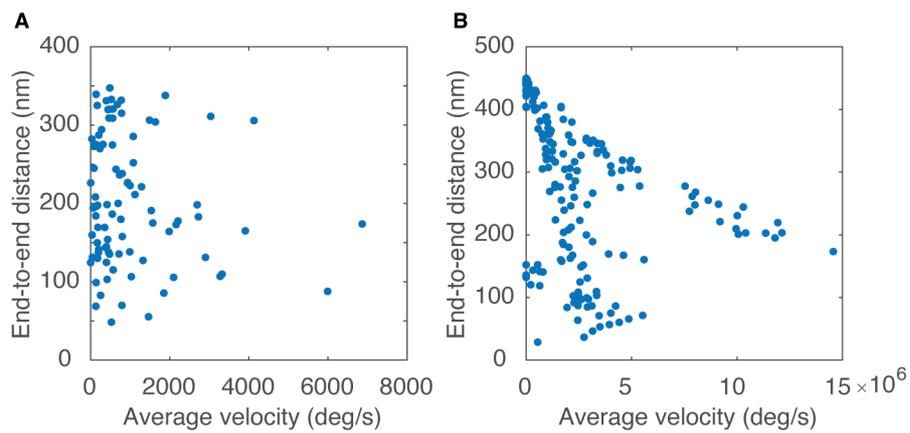

**Fig. S10. Scatter plot of the end-to-end distance of the rotors vs their rotation speed from experiment (A) and simulation (B).**

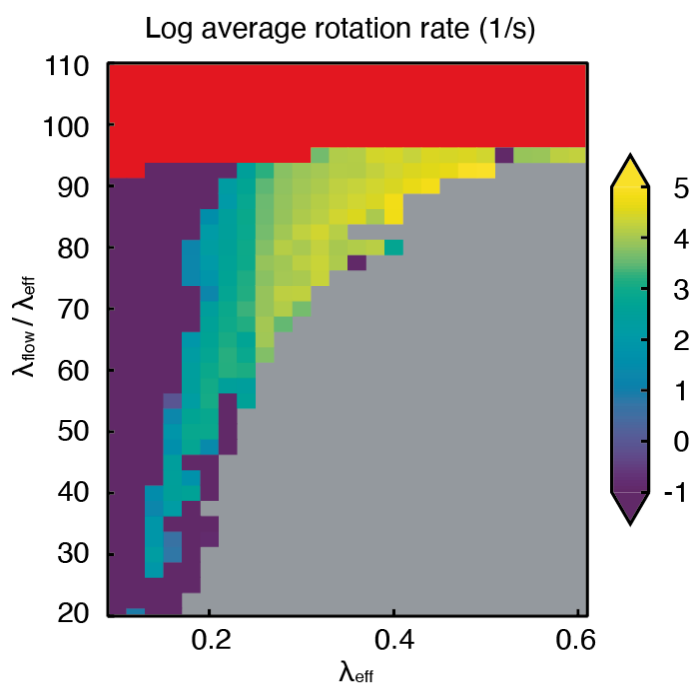

**Fig. S11. Phase diagram of observed behaviors in simulated 6hbs.**
The figure displays the results of a parameter scan where the strength of the electrophoresis and the electroosmotic drag are varied. Color indicates the base-10 logarithm of the rotation rate averaged over different initial rod placements. All data in purple/green/yellow exhibited sustained rotations. Parameters that did not yield any stationary states are marked separately, viz., as non-docking (red) or translocating (grey) 6hbs. Stably rotating states with a finite rotation rate are observed in the intermediate regime where all forces allow for a non-trivial balance. The ratio of the flow factor ($\lambda_{flow}$) and EP mobility ($\lambda_{eff}$) is given in the Y axis. The X axis describes the strength of the electrophoretic drift towards the pore, with the value of 1 representing the bare-charge electrostatic force. An experimental scan that varies the electric field strength is equivalent to a horizontal line through the phase diagram, which, from low to high values, would yield a docked but nonrotating rotor, a rotor that rotates with increasing speeds for higher fields, and finally translocation of the 6hb through the nanopore – all in agreement with the experimental observations.

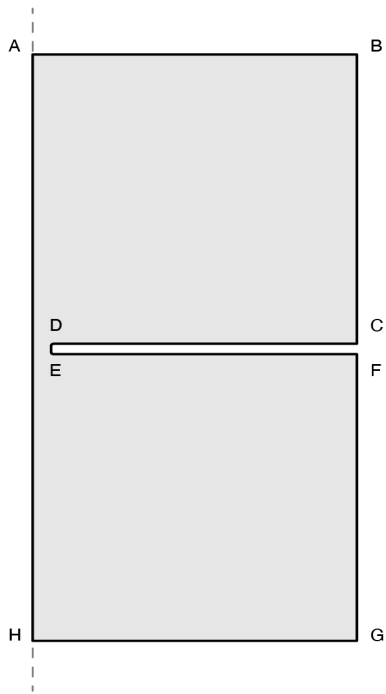

**Fig. S12. Geometry used in simulating electroosmotic flow through the nanopore.** ABHG denotes an XZ plane across the nanopore axis AH for a nanopore that is located midway between A and H. Rotation symmetry defines the electroosmotic flow in 3D by rotating this plane about the AH axis.

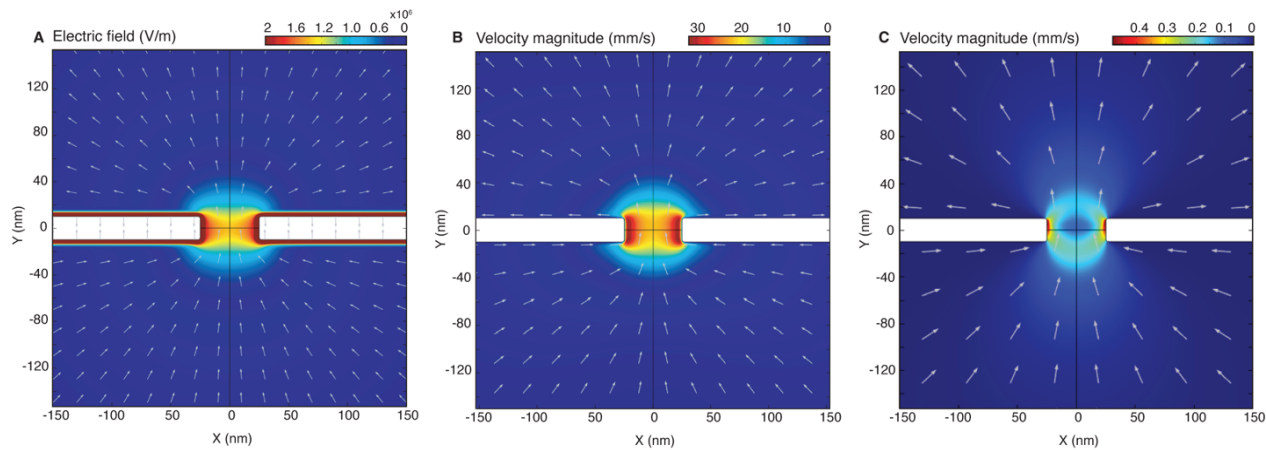

**Fig. S13. Simulated electroosmotic and diffusioosmotic flow through the nanopore.**

(A) Simulated electric field under 100 mV bias voltage and 50 mM NaCl. (B) Simulated electroosmotic flow distribution under 100 mV bias voltage and 50 mM NaCl. (C) Simulated diffusioosmotic flow distribution under 550 mM:50 mM NaCl gradient."

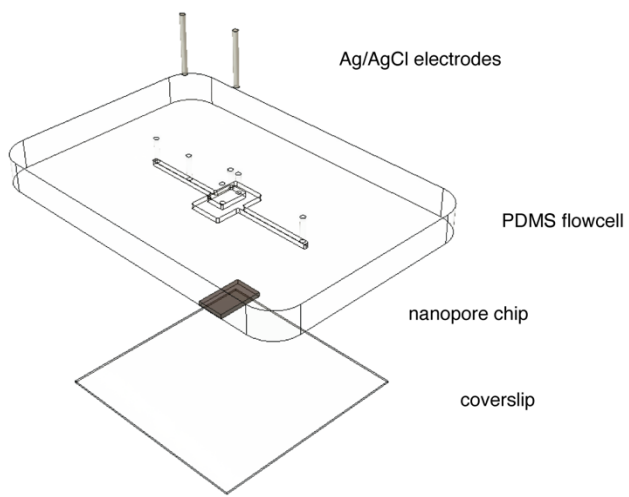

**Fig. S14. Schematic of the PDMS flow cell used in the experiments.**

**Table S1. Variables used in the simulations**

| Name | Value | Description |
| --- | --- | --- |
| W | 500 [nm] | width of simulation box |
| H | 1000 [nm] | height of simulation box |
| L | 20 [nm] | thickness of the membrane |
| R | 25 [nm] | radius of the nanopore |
| sigma | -0.1x e [nm-2] | surface charge density [e/nm2] |
| epsilon_water | 78 | permittivity of water |
| e | 1.602176634e-19 [C] | elementary charge |
| $V_{bias}$ | 100 [mV] | bias voltage |
| $c_{Na}$ | 0.05 [mol/L] | salt concentration |
| $c_{Cl}$ | 0.05 [mol/L] | salt concentration |
| $D_{Cl}$ | 2.03e-9 [m2/s] | diffusion coefficient, Cl- |
| $D_{Na}$ | 1.33e-9 [m2/s] | diffusion coefficient, Na+ |
| F | 96485.33212 [C/mol] | Faraday Constant |
| T | 300 [K] | temperature |

**Table S2. Geometry settings**

| Domain/Boundary | Settings |
|---|---|
| AB | W |
| AH | H |
| DE | L |
| CD | W-R |

**Table S3. Electrostatics (es)**

| Domain/Boundary | Settings |
| --- | --- |
| AH | Axial Asymmetry |
| BC, FG | Zero Charge |
| AB | Ground: $V = 0$ |
| HG | Electric Potential: $V = V_{bias}$ |
| CDEF | Surface Charge Density: sigma |
| ABCDEFGH | Space Charge Density: $F \times (Na - Cl)$ [C/m$^3$] |

**Table S4. Creeping flow (spf)**

| Domain/Boundary | Settings |
| --- | --- |
| AH | Axial Asymmetry |
| ABCDEFGH | Volume force, $F_r = (Na - Cl) \times F \times es.E_r,$ $F_z = (Na - Cl) \times F \times es.E_z,$ |
| AB, HG | Open boundary, Normal stress $f = 0$ [N/m2] |
| BC, FG | Wall, slip |
| CDEF | Wall, no slip |

**Table S5. Transport of Diluted Species (tds)**

| Domain/Boundary | Settings |
|---|---|
| AH | Axial Asymmetry |
| ABCDEFGH | Velocity Field (spf) |
| | Concentration ($Na^+$) = Na, Concentration ($Cl^-$)=Cl |
| | Diffusion coefficient: $D_{Na}$, $D_{Cl}$ |
| | Electric Potential (es) |
| | Migration in Electric Field: V(es), Nernst-Einstein relation |
| | Charge Number: $Z_{Na} = 1, Z_{Cl} = -1$ |
| | Initial Values: $Na = c_{Na}, Cl = c_{Cl}$ |
| BCDEFG | No flux |
| AB, HG | Concentration: $Na = c_{Na}, Cl = c_{Cl}$ |

**Movie S1: Rotary motion of DNA rotors on nanopores.** Top row: Raw video of the Cy5 channel of each rotor. Middle row: Corresponding single-particle localization results of both ends of the DNA rotors. The position of the current frame is marked as orange and blue dots, and the trajectory of the 10 frames before the current frame is shown as solid lines. The two dots are connected with a red bar. Bottom row: Corresponding cumulative angular displacement $\theta(t)$. All plots in the video are synced. The video playback frame rate is 40 fps, which is around 10 times slower than the original data (450-500 fps).

**Movie S2: Simulated DNA rotors on nanopores.** Top view of a collection of different simulated DNA rotors. The 6hb rods are shown in orange and the rim of the pore in red. All parameters were the same for each panel in this video, except for the initial placement of the 6hb rod (shown in blue) on the nanopore. Simulations that terminated early due to translocation of the rod through the pore are marked by a grey background.

**Movie S3-S6: Examples of different bending configurations in 3D simulations of DNA rods on nanopores.** In each video, a portion of the membrane is shown in grey, the rim of the pore is highlighted in red, and a 3D rendering of the motion of the DNA rod is displayed.

**Additional references**